# LabView® Interface for School-Network DAQ Card

Hans Berns, T. H. Burnett, Richard Gran, Graham Wheel, R. Jeffrey Wilkes, *Univ. of Washington, Seattle, WA*

Dan Claes, Jared Kite, Gregory Snow, *Univ.of Nebraska, Lincoln, NE*

*Abstract*— **As described elsewhere at this conference, a low-cost DAQ card has been developed for school-network cosmic ray detector projects, providing digitized data from photomultiplier tubes via a standard serial interface. To facilitate analysis of these data and to provide students with a starting point for custom readout systems, a model interface has been developed using the National Instruments LabVIEW® system. This user-friendly interface allows one to initialize the trigger coincidence conditions for data-taking runs and to monitor incoming or pre-recorded data sets with updating singles- and coincidence-rate plots and other user-selectable histograms.**

## I. INTRODUCTION

THE WALTA [1] project aims to create large-area extensive air shower (EAS) detector array for ultra-high energy (UHE) cosmic rays, by installing mini-arrays of scintillation counter detectors in secondary schools, in the Seattle, WA area. Data taken at individual school sites are shared via Internet connections and searched for multi-site coincidence events.

WALTA collaborated with CROP [2], a similar project based at the U. of Nebraska at Lincoln, to salvage plastic scintillators, photomultiplier tubes (PMTs) and associated high voltage supplies from surplus equipment at the CASA detector [3] site at Dugway, UT.

Individual detector stations each consist of 4 scintillation counter modules, front-end electronics, and a GPS receiver, as shown in Fig. 1. Preliminary training of secondary school teachers and students was conducted using obsolete NIM crates and fast electronics (discriminators, coincidence and scaler modules) loaned from Fermilab. These modules are now being replaced by the new DAQ cards, which add GPS timing and a simple RS232 computer interface.

The QuarkNet [4] DAQ card provides a low-cost alternative to standard particle and nuclear physics fast pulse electronics modules. The board, which can be produced at a cost of less than US$500, produces trigger time and pulse edge time data for 2 to 4-fold coincidence levels, via a universal RS-232 serial port interface, usable with any PC. Details of the DAQ card have been presented elsewhere [5].

The DAQ cards produce a stream of data in the form of ASCII text strings, in distinct formats for different functions. However, the highly compact data format, using hexadecimal coded data items without user-friendly prompts, makes use of the card via a simple terminal window rather challenging for typical users.

The user's host computer, a desktop PC which may have a Windows, Mac or Linux OS, is connected to the DAQ card via a standard RS232 serial cable. For schools that have relatively new Macintosh models, with USB ports only, a commercial USB-to-RS232 adapter can be inserted. The board and GPS unit are powered by a conventional 5VDC, 800 mA power adapter of the type available at consumer electronics stores.

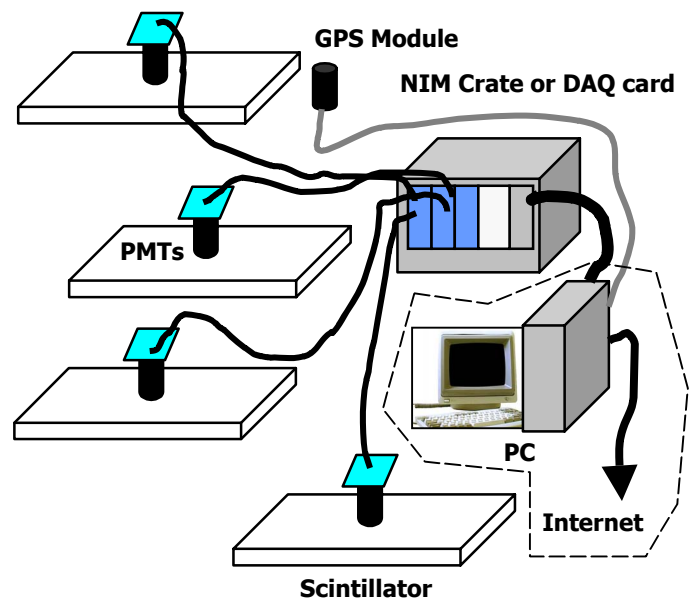

Fig. 1. School-network air shower detector stations. CROP and WALTA use PMTs and counters salvaged from the CASA experiment. Initial installations used NIM crates borrowed from Fermilab PREP, which will be replaced by the DAQ cards described in this paper.

Manuscript received October 29, 2003. This work was supported in part by the U.S. Department of Energy, Quarknet, and the U.S. National Science Foundation. Contact: wilkes@u.washington.edu.

The simplest way to operate the board is by opening a terminal window linked to the PC's serial port. Table I shows the help screen displayed upon board startup, listing the functions implemented. Fig. 2 shows a typical terminal screen during data-taking.

Commands can be directly entered via the keyboard, and the resulting output viewed (and if desired, captured to a log file) from the screen. Typically, the user would enter commands to define the trigger logic level (1- to 4-fold majority), and enable counting. At the end of the desired time interval, counting can be disabled and the trigger count read out. For each trigger, the display shows one or more lines of hexadecimal encoded data, providing time of trigger, pulse leading and trailing edge times relative to the trigger time, and information needed to determine the GPS time of the trigger to 24 ns precision. Other commands can be used to directly interrogate the GPS module, or act upon the counter registers, etc.

```
Scintillator Card, QNET2  Firmware Ver2.3, 09/09/03   HE=Help
Serial#=1002   uC_Volts=3.3   uC_TempC=26.6   GPS_TempC=235.0   kPa=0

Barometer - BA=Display, BA bb.b gg.g calibrates kPa Baseline, Gain (See HB).
Counter   - CE=Enable, CD=Disable, Controls TMC Running bit @ CPLD CCR1.
DC        - Display Counters and Control Registers of CPLD, address 0-4.
DF        - Display Scalar Fifo Data (first 12 Bytes as three 32bit numbers).
DG        - Display GPS Date, Time, Position and Status.
DS        - Display Scalar Fifo, Counters 0-3, Triggers, and 1_PPS Time.
DT        - Display Time Control Registers of TMC, address 0-3.
Flash     - FL=Load Binary File, FR=Read SumCheck, FC=Copy_to_CPLD.
GP        - Init Link with GPS unit (GGA=1/sec, RMC=1/sec, disable others).
Help      - HF=Trigger format, HS=Status format, HB=Barometer format.
NA n      - NMEA GPS Data Append (n==1 On),(n!=1 Off), add GPS to output.
NM n      - NMEA GPS Data Echo (n==1 On),(n!=1 Off), (GPS_Baud=9600).
Reset     - RB=TMC+CPLD,  RE=MSP430+TMC+CPLD.
SA n      - SA=Save TMC+CPLD Registers to Flash, (SA 1=Restore Defaults).
SB n      - Set Baud Rate (PC Link), 1=19200, 2=38400, 3=57600, 4=115200.
SN nnnn   - Serial Number(BCD), SN=Display Number, SN nnnn=Store Serial
Number.
ST n      - Send Status Data (n==1 On),(n!=1 Off), (See HF).
TH        - Thermometer Data Display, -40 to 99 degrees C.
WC mm nn  - Write Counter Control Registers CPLD address mm with data nn.
WT mm nn  - Write Time Control Registers TMC address mm with data nn.
```

The QuarkNet LabView® interface software allows the user to send setup and operating commands to the card, and to log data of various types. Log files can be output in MS Excel-compatible format, for direct handling by users who do not have programming skills.

The GUI is divided into pages with different functions, for setup, housekeeping data, routine data-taking, etc. Figure 3 shows the configuration window, which can be opened from any tab of the GUI. Figures 4 through 6 show examples of the console, GPS record [7] and event timing windows.

```
>BA
2480 0          ← Request barometric pressure
>TH
235.1           ← Request temperature at card
>WC 04 FF       ← Write to register 4: enable all channels
@0004=003F
>DS             ← Start collecting trigger data…
@00 000016EA
@01 00001EC2    ← card first reports its counter contents
@02 000067E3
@03 00000000
@04 0024E1E4
@05 B25EBE0E               (…then data stream)
>B388F1B7 AE 01 31 01 3E 01 00 01 B25EBE0E 195439.340 060603 V 00 2 +0638
B388F1B8 01 00 01 37 32 2A 00 01 B25EBE0E 195439.340 060603 V 00 2 +0638
B388F1B9 27 22 00 01 00 01 00 01 B25EBE0E 195440.340 060603 V 00 2 -0361
B388F1BA 00 01 31 34 00 01 00 01 B25EBE0E 195440.340 060603 V 00 0 -0361
(etc)
```

Fig. 2. Typical terminal window, showing raw data stream from the QuarkNet DAQ card. Commands and responses shown illustrate reading the on-board barometer and temperature sensor, reading scaler contents, and the data stream produced while logging coincidences. These data serve as input to the LabView module.

Users may also prepare a script to implement a sequence of commands, or compile custom software to operate the card directly.

## II. THE WALTA LABVIEW INTERFACE

We have developed an interface to the QuarkNet DAQ card using the National Instruments LabView® environment [6]. This interfacing environment is available at low cost to educational institutions, has a very large user base worldwide, and provides a simple graphical user toolkit with which it is possible to construct highly capable real-time programs. The large base of applications and program elements makes it easy for users to acquire help and training.

TABLE I
DAQ CARD COMMANDS (HELP SCREEN DISPLAY)

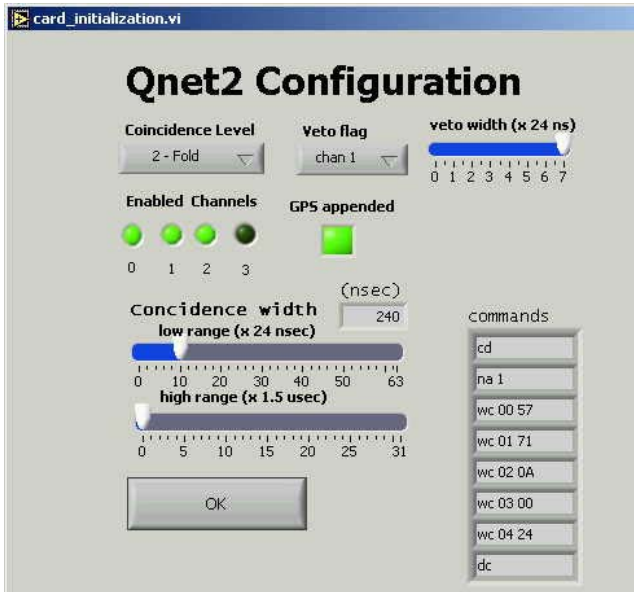

Fig. 3. DAQ card configuration window..

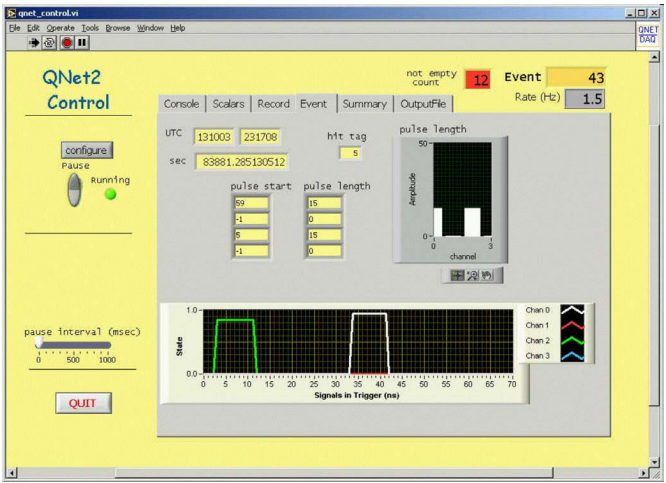

Fig. 4. Event timing data window.

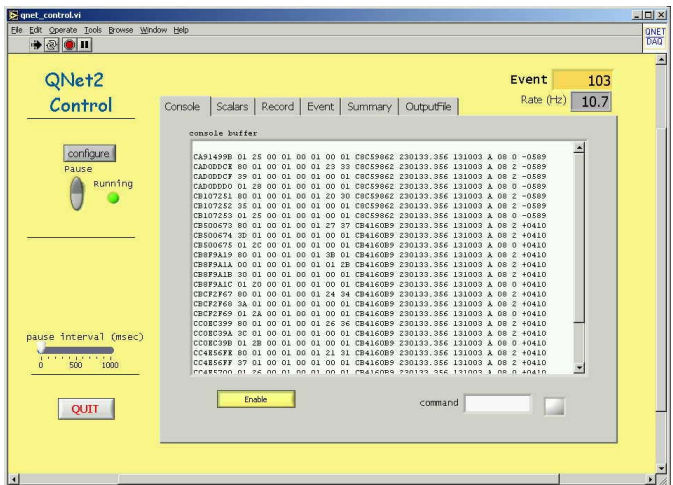

Fig. 5. "Console" data tab, showing raw hexadecimal data stream and allowing manual input of commands.

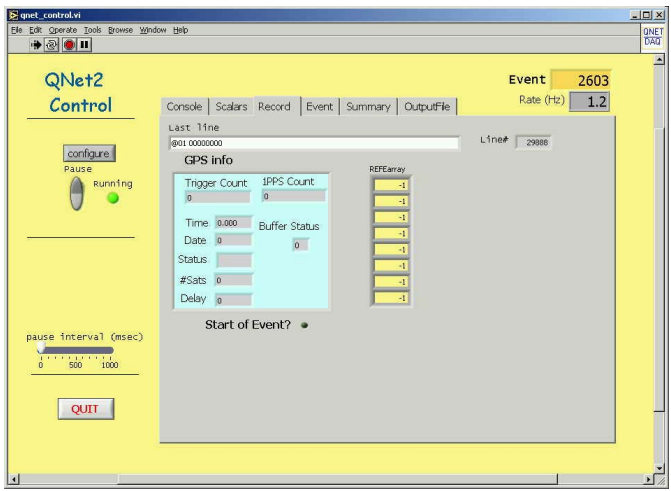

Fig. 6. GPS record data tab.

## III. FUTURE PLANS

We are currently refining the software described, with high school groups as well as UW undergraduates serving as beta testers. We anticipate that participating students and teachers will contribute to the further development of this software.

## IV. ACKNOWLEDGMENT

We wish to thank Mark Buchli, Paul Edmon, Ben Laughlin, and Jeremy Sandler for valuable assistance in testing and debugging the boards, and the CROP and WALTA teachers for their patience with the development process. Special thanks are due to Tom Jordan, Director of the QuarkNet program, for his support.